\documentclass[a4paper,11pt]{article}

\usepackage{fullpage} % Package to use full page
\usepackage{parskip} % Package to tweak paragraph skipping
\usepackage{amsmath}
\usepackage{hyperref}
\usepackage{amsmath,amsfonts,amsthm} % Math packages
\usepackage{graphicx}
\usepackage{listings}
\usepackage{caption}
\usepackage{color}
\usepackage{float}
\definecolor{codegreen}{rgb}{0,0.6,0}
\definecolor{codegray}{rgb}{0.5,0.5,0.5}
\definecolor{codepurple}{rgb}{0.58,0,0.82}
\definecolor{backcolour}{rgb}{0.95,0.95,0.92}
\definecolor{brown}{rgb}{0.59, 0.29, 0.0}
\definecolor{beaublue}{rgb}{0.74, 0.83, 0.9}
\definecolor{orange}{rgb}{1.0, 0.5, 0.0}
\definecolor{darkslategray}{rgb}{0.18, 0.31, 0.31}

\def\XXint#1#2#3{{\setbox0=\hbox{$#1{#2#3}{\int}$}
		\vcenter{\hbox{$#2#3$}}\kern-.5\wd0}}

\makeatletter
\let\oldabs\abs
\def\abs{\@ifstar{\oldabs}{\oldabs*}}
\let\oldnorm\norm
\def\norm{\@ifstar{\oldnorm}{\oldnorm*}}
\makeatother
\definecolor{keywords}{RGB}{255,0,90}
\definecolor{comments}{RGB}{0,0,113}
\definecolor{red}{RGB}{160,0,0}
\definecolor{green}{RGB}{0,150,0}
\definecolor{codegreen}{rgb}{0,0.6,0}
\definecolor{codegray}{rgb}{0.5,0.5,0.5}
\definecolor{codepurple}{rgb}{0.58,0,0.82}
\definecolor{backcolour}{rgb}{0.95,0.95,0.92}
\definecolor{brown}{rgb}{0.59, 0.29, 0.0}
\definecolor{beaublue}{rgb}{0.74, 0.83, 0.9}
\definecolor{orange}{rgb}{1.0, 0.5, 0.0}
\definecolor{darkslategray}{rgb}{0.18, 0.31, 0.31}
\definecolor{deepblue}{rgb}{0,0,0.5}
\definecolor{deepred}{rgb}{0.6,0,0}
\definecolor{deepgreen}{rgb}{0,0.5,0}
\lstdefinestyle{myMatlabstyle}{
	language=Matlab,
	backgroundcolor=\color{white},   
	commentstyle=\color{codegreen},
	keywordstyle=\color{blue},
	identifierstyle=\color{brown},
	numberstyle=\tiny\color{codegray},
	stringstyle=\color{orange},
	basicstyle=\footnotesize,
	breakatwhitespace=false,         
	breaklines=true,                 
	captionpos=b,                    
	keepspaces=true,                
	numbers=left,                    
	numbersep=5pt,                  
	showspaces=false,                
	showstringspaces=false,
	showtabs=false,                  
	tabsize=2
}
\lstdefinestyle{myPythonstyle}{
	language=Python, 
	basicstyle=\ttfamily\small, 
	keywordstyle=\color{blue},
	commentstyle=\color{green},
	stringstyle=\color{red},
	showstringspaces=false,
	identifierstyle=\color{black},
}
\usepackage{tikz}
\usepackage{array}
\usepackage{mdwmath}
\usepackage{mdwtab}
\newcommand{\footremember}[2]{%
    \footnote{#2}
    \newcounter{#1}
    \setcounter{#1}{\value{footnote}}%
}
\usepackage[caption=false,font=footnotesize]{subfig}
\title{\vspace{-2.5cm} The right way to teach the FFT}
\author{Jithin D. George \footremember{alley}{Department of Applied Mathematics, University of
Washington, Seattle, WA (jithindgeorge93@gmail.com)}}
\date{20 May 2018}
\newenvironment{mat}{\left[ \begin{array}{ccccccccccccc}}{\end{array}\right]}
\newcommand\bcm{\begin{mat}}
	\newcommand\ecm{\end{mat}}
\AtBeginDocument{\hypersetup{pdfborder={0 0 0}}}% <-----------

\begin{document}
\maketitle

\begin{abstract}
The algorithm behind the Fast Fourier Transform has a simple yet beautiful 
geometric interpretation that is often lost in translation in a classroom. 
This article provides a visual perspective
which aims to capture the essence of it.
\end{abstract}

Students are often confused when they encounter the Fast Fourier Transform (FFT) for the first time. The author believes the confusion stems from two sources.

\begin{enumerate}
 \item 
  The belief that one needs to understand the Fourier transform completely to understand the FFT. This is not true. The FFT simply is an efficient way of computing sums of a special form and the terms in the Discrete Fourier Transform (DFT) \eqref{fft} just happened to be in that form.

  \begin{equation}\label{fft}
A_k = \sum_{n=0}^{N-1} a_n e^{- i  \frac{2\pi n}{N}k}
\end{equation} 

  \item
  The Cooley-Tukey Algorithm \cite{key-1}. This is the heart of the FFT. The idea is that it is possible to decompose the DFT of a sequence of terms into the DFT of the even terms and the DFT of the odd terms. When applied recursively, this results in a computational cost of $O(N\log{}N)$. The following decomposition of $A_k$ into odd and even terms is generally used to illustrate the idea.
  \begin{equation}\label{CT}
 \sum_{n=0}^{N-1} a_n e^{- i  \frac{2\pi n}{N}k}= \sum_{n=0}^{N/2-1} a_{2n} e^{- i  \frac{2\pi (2n)}{N}k} + \sum_{n=0}^{N/2-1} a_{2n+1} e^{- i  \frac{2\pi (2n+1)}{N}k}
\end{equation} 
\end{enumerate}

Let us take a simplified look at the terms in a DFT.

  \begin{equation}\label{fft2}
A_k = \sum_{n=0}^{N-1} a_n e^{- i  n\frac{2\pi}{N}k}= \sum_{n=0}^{N-1} a_n e^{i k \theta_n} 
\end{equation} 

$a_n e^{i k \theta_n}$ can be visualized as the value $a_n$ located at an angle of $k\theta_n$ on a unit circle in the complex plane. 
As n goes from 0 to N-1, the $\theta_n$s divide the circle into N arcs of angle $\frac{2\pi}{N}$. Each term in the summation in \eqref{fft2} is a multiple of a point on the unit circle in the complex plane.

\begin{tikzpicture}
 \def \n {8}
\def \radius {1.02}
\foreach \s in {0,1,2,3,4,5,6,7}
{
  \node[circle,fill,radius = 0.2 pt , inner sep= 0.05pt,label=above:$a_\s$] at ({7+\radius*cos(360/\n * (\s ))}, {\radius*sin(360/\n * (\s ))} ) {$\s$};
}

\node[anchor=base] at (4, 0)
                {$=$};
\node[anchor=base] at (0, 0)
                {$A_1\{a_0,a_1,a_2,a_3,a_4,a_5,a_6,a_7\}$};  
                
                \draw (7,0) circle (\radius);
\end{tikzpicture}

 With this geometric view, the Cooley-Tukey algorithm in \eqref{CT} becomes obvious through the diagram below.

\newpage

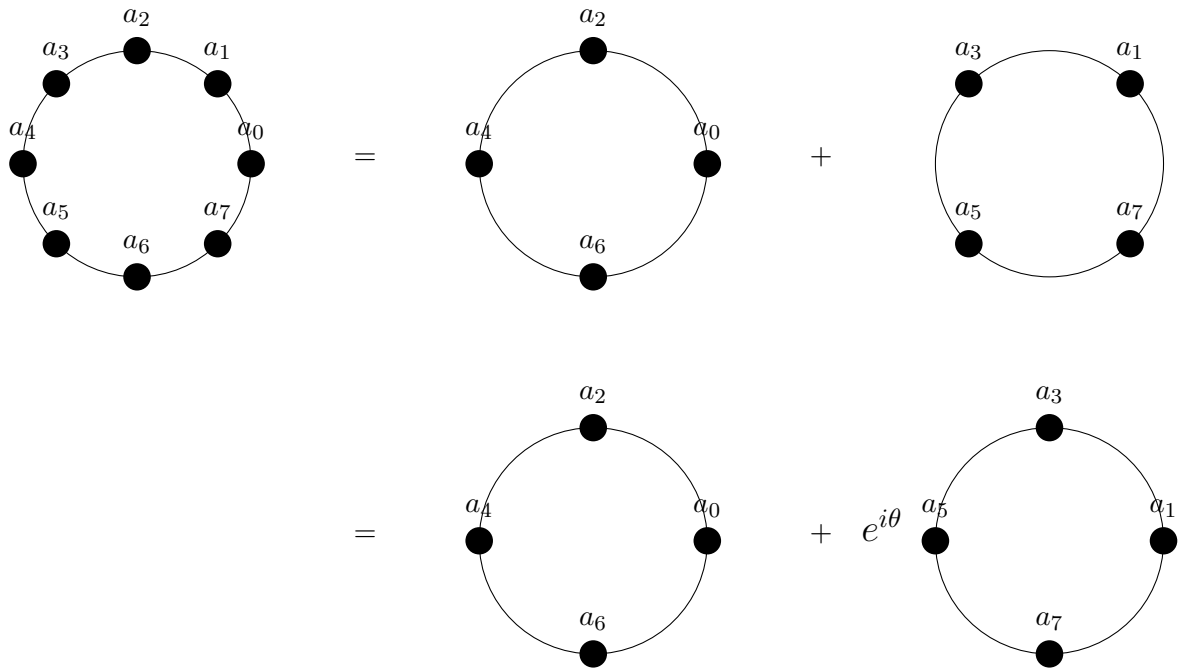
\begin{tikzpicture}

\def \n {8}
\def \radius {1.5}
\def \margin {8}
\def \cem{8}
\def \cer{14}
\def \ced{2}
\def \c_32{1}
\def \down{5}

\foreach \s in {0,1,2,3,4,5,6,7}
{
  \node[circle,fill,radius = 0.2 pt , inner sep= 0.5pt,label=above:$a_\s$] at ({\ced+\radius*cos(360/\n * (\s ))}, {\ced+\radius*sin(360/\n * (\s ))} ) {$\s$};
}
\foreach \s in {0,2,4,6}
{
  \node[circle,fill,radius = 0.2 pt , inner sep= 0.5pt,label=above:$a_\s$] at ({\cem+\radius*cos(360/\n * (\s ))}, {\ced+\radius*sin(360/\n * (\s ))} ) {$\s$};
}

\foreach \s in {1,3,5,7}
{
  \node[circle,fill,radius = 0.2 pt , inner sep= 0.5pt,label=above:$a_\s$] at ({\cer+\radius*cos(360/\n * (\s ))}, {\ced+\radius*sin(360/\n * (\s )) } ) {$\s$};
}
\draw (\ced,\ced) circle (\radius);
\draw (\cem,\ced) circle (\radius);
\draw (\cer,\ced) circle (\radius);
%\draw (0,0) -- (4,0);

\node[anchor=base] at (5, \ced )
                {$=$};
\node[anchor=base] at (11, \ced )
                {$+$}; 
                
\node[anchor=base] at (5, \ced -\down)
                {$=$};
\node[anchor=base] at (11, \ced -\down)
                {$+$};  
     \node[anchor=base] at (11.8, \ced -\down)
                {\Large $e^{i\theta}$}; 
                
  \foreach \s in {0,2,4,6}
{
  \node[circle,fill,radius = 0.2 pt , inner sep= 0.5pt,label=above:$a_\s$] at ({\cem+\radius*cos(360/\n * (\s ))}, {\ced-\down+\radius*sin(360/\n * (\s ))} ) {$\s$};
}

\foreach \s in {1,3,5,7}
{
  \node[circle,fill,radius = 0.2 pt , inner sep= 0.5pt,label=above:$a_\s$] at ({\cer+\radius*cos(360/\n * (\s - 1))}, {\ced-\down+\radius*sin(360/\n * (\s - 1)) } ) {$\s$};
}
\draw (\cer,\ced-\down) circle (\radius);
\draw (\cem,\ced-\down) circle (\radius);
\end{tikzpicture}

We are able to compute sums like the FFT in this way because the odd terms are a 'rotation' away from the even terms. This is quite elegant but it in itself, doesn't provide any new computational efficiency. We decompose a sum into two smaller sums of half the size. But we still have to do all the sums either way. What gives the FFT its computational efficiency is the fact that the smaller sums can be `recycled' to get new sums.

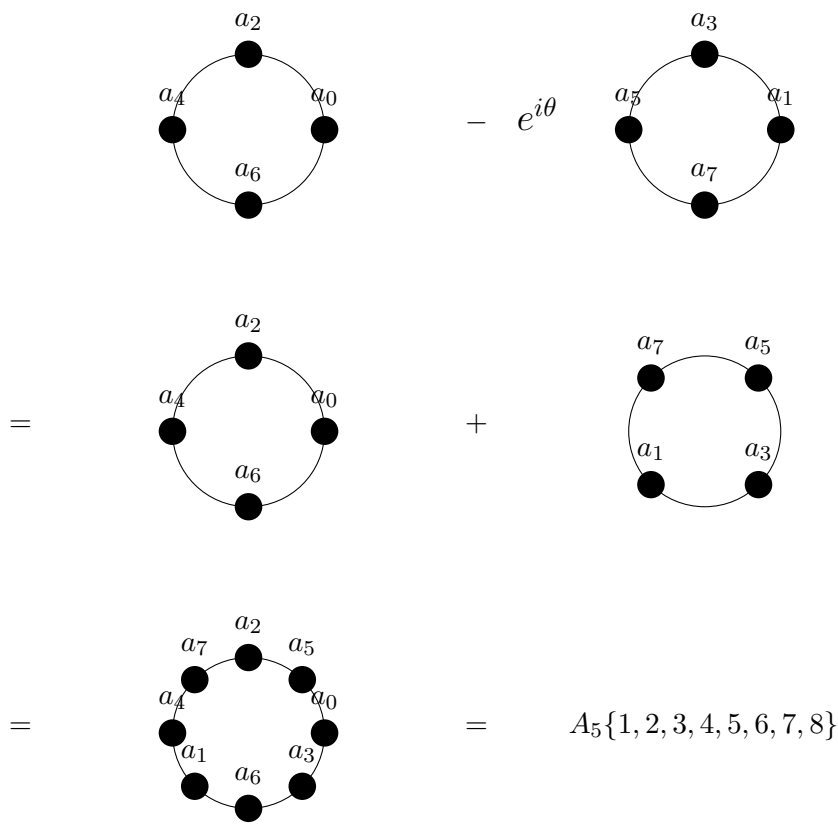
\begin{tikzpicture}

\def \n {8}
\def \radius {1}
\def \margin {8}
\def \cem{8}
\def \cer{14}
\def \ced{2}
\def \c_32{1}
\def \down{4}

\foreach \s in {0,2,4,6}
{
  \node[circle,fill,radius = 0.2 pt , inner sep= 0.5pt,label=above:$a_\s$] at ({\cem+\radius*cos(360/\n * (\s ))}, {\ced+\radius*sin(360/\n * (\s ))} ) {$\s$};
}

\foreach \s in {1,3,5,7}
{
  \node[circle,fill,radius = 0.2 pt , inner sep= 0.5pt,label=above:$a_\s$] at ({\cer+\radius*cos(5*360/\n * (\s ))}, {\ced-\down+\radius*sin(5*360/\n * (\s )) } ) {$\s$};
}

\draw (\cem,\ced) circle (\radius);
\draw (\cer,\ced) circle (\radius);
%\draw (0,0) -- (4,0);

\node[anchor=base] at (11, \ced -\down)
                {$+$}; 
                
\node[anchor=base] at (5, \ced - \down)
                {$=$};

\node[anchor=base] at (11, \ced )
                {$-$};  
     \node[anchor=base] at (11.8, \ced)
                {\Large $e^{i\theta}$}; 
                
  \foreach \s in {0,2,4,6}
{
  \node[circle,fill,radius = 0.2 pt , inner sep= 0.5pt,label=above:$a_\s$] at ({\cem+\radius*cos(360/\n * (\s ))}, {\ced-\down+\radius*sin(360/\n * (\s ))} ) {$\s$};
}

\foreach \s in {1,3,5,7}
{
  \node[circle,fill,radius = 0.2 pt , inner sep= 0.5pt,label=above:$a_\s$] at ({\cer+\radius*cos(360/\n * (\s - 1))}, {\ced+\radius*sin(360/\n * (\s - 1)) } ) {$\s$};
}
\draw (\cer,\ced-\down) circle (\radius);
\draw (\cem,\ced-\down) circle (\radius);
\draw (\cem,\ced-2*\down) circle (\radius);

\node[anchor=base] at (5, \ced - 2*\down)
                {$=$};
 \node[anchor=base] at (11, \ced - 2*\down)
                {$=$}; 
                
  \node[anchor=base] at (\cer , \ced - 2*\down)
                {$A_5\{1,2,3,4,5,6,7,8\}$};                 
 \foreach \s in {0,1,2,3,4,5,6,7}
{
  \node[circle,fill,radius = 0.2 pt , inner sep= 0.5pt,label=above:$a_\s$] at ({\cem+\radius*cos(5*360/\n * (\s ))}, {\ced-2*\down+\radius*sin(5*360/\n * (\s ))} ) {$\s$};
}               
                \end{tikzpicture}
                
The two terms which when added give $A_1$ can be `recycled' by subtracting them to give $A_5$. This definitely has to save some computational cost. How much ? To get the fine details, we would have to work out a simple example.

\textbf{An example}

Let's have a look at the DFT of the vector $\{a_0,a_1,a_2,a_3\}$.

\begin{tikzpicture}

\def \n {8}
\def \radius {1}
\def \margin {8}
\def \cem{8}
\def \cer{14}
\def \ced{5}
\def \c_32{1}
\def \down{3}

\foreach \s in {1}
{
  \node[circle,fill,radius = 0.2 pt , inner sep= 0.5pt,label=right:$a_0\,a_\s\,a_2\,a_3$] at ({\ced+\radius*cos(360/\n * (\s - 1))}, {\ced+\radius*sin(360/\n * (\s - 1))} ) {$\s$};
}

\foreach \s in {1}
{
  \node[circle,fill,radius = 0.2 pt , inner sep= 0.5pt,label=right:$a_0\,a_2$] at ({\ced+\radius*cos(360/\n * (\s - 1))}, {\ced-\down+\radius*sin(360/\n * (\s - 1))} ) {$\s$};
}
\foreach \s in {1}
{
  \node[circle,fill,radius = 0.2 pt , inner sep= 0.5pt,label=right:$a_1\,a_3$] at ({\ced+5+\radius*cos(360/\n * (\s - 1))}, {\ced-\down+\radius*sin(360/\n * (\s - 1))} ) {$\s$};
}

\node[anchor=base] at (2, \ced )
                {$A_0$};
          \node[anchor=base] at (3, \ced )
                {$=$};
          \node[anchor=base] at (3, \ced-\down  )
                {$=$};                
\draw (\ced,\ced) circle (\radius);
\draw (\ced,\ced-\down) circle (\radius);
\draw (\ced+5,\ced-\down) circle (\radius);
%\draw (0,0) -- (4,0);

\node[anchor=base] at (8, \ced -\down)
                {$+$};

                \end{tikzpicture}
                
\begin{tikzpicture}

\def \n {4}
\def \radius {1}
\def \margin {8}
\def \cem{8}
\def \cer{14}
\def \ced{5}
\def \c_32{1}
\def \down{3}

\foreach \s in {0,1,2,3}
{
  \node[circle,fill,radius = 0.2 pt , inner sep= 0.5pt,label=above:$a_\s$] at ({\ced+\radius*cos(360/\n * (\s ))}, {\ced+\radius*sin(360/\n * (\s ))} ) {$\s$};
}

\foreach \s in {0,2}
{
  \node[circle,fill,radius = 0.2 pt , inner sep= 0.5pt,label=above:$a_\s$] at ({\ced+\radius*cos(360/\n * (\s ))}, {\ced-\down+\radius*sin(360/\n * (\s ))} ) {$\s$};
}
\foreach \s in {1,3}
{
  \node[circle,fill,radius = 0.2 pt , inner sep= 0.5pt,label=right:$a_\s$] at ({\ced+5+\radius*cos(360/\n * (\s - 1))}, {\ced-\down+\radius*sin(360/\n * (\s - 1))} ) {$\s$};
}

\node[anchor=base] at (2, \ced )
                {$A_1$};
          \node[anchor=base] at (3, \ced )
                {$=$};
          \node[anchor=base] at (3, \ced-\down  )
                {$=$};                
\draw (\ced,\ced) circle (\radius);
\draw (\ced,\ced-\down) circle (\radius);
\draw (\ced+5,\ced-\down) circle (\radius);
%\draw (0,0) -- (4,0);

\node[anchor=base] at (7, \ced -\down)
                {$+$}; 
                
\node[anchor=base] at (8, \ced -\down)
                {$e^{i\theta}$};

                \end{tikzpicture}
       
       \begin{tikzpicture}

\def \n {4}
\def \radius {1}
\def \margin {8}
\def \cem{8}
\def \cer{14}
\def \ced{5}
\def \c_32{1}
\def \down{3}

\foreach \s in {0}
{
  \node[circle,fill,radius = 0.2 pt , inner sep= 0.5pt,label=right:$a_\s\, a_{2}$] at ({\ced+\radius*cos(2*360/\n * (\s ))}, {\ced+\radius*sin(2*360/\n * (\s ))} ) {$\s$};
}

\foreach \s in {1}
{
  \node[circle,fill,radius = 0.2 pt , inner sep= 0.5pt,label=right:$a_\s\, a_{3}$] at ({\ced+\radius*cos(2*360/\n * (\s ))}, {\ced+\radius*sin(2*360/\n * (\s ))} ) {$\s$};
}
\foreach \s in {0}
{
  \node[circle,fill,radius = 0.2 pt , inner sep= 0.5pt,label=right:$a_\s\,a_2$] at ({\ced+\radius*cos(2*360/\n * (\s ))}, {\ced-\down+\radius*sin(2*360/\n * (\s ))} ) {$\s$};
}
\foreach \s in {1}
{
  \node[circle,fill,radius = 0.2 pt , inner sep= 0.5pt,label=right:$a_\s\,a_3$] at ({\ced+5+\radius*cos(360/\n * (\s - 1))}, {\ced-\down+\radius*sin(360/\n * (\s - 1))} ) {$\s$};
}

\node[anchor=base] at (2, \ced )
                {$A_2$};
          \node[anchor=base] at (3, \ced )
                {$=$};
          \node[anchor=base] at (3, \ced-\down  )
                {$=$};                
\draw (\ced,\ced) circle (\radius);
\draw (\ced,\ced-\down) circle (\radius);
\draw (\ced+5,\ced-\down) circle (\radius);
%\draw (0,0) -- (4,0);

\node[anchor=base] at (8, \ced -\down)
                {$-$};

                \end{tikzpicture}
        
       \begin{tikzpicture}

\def \n {4}
\def \radius {1}
\def \margin {8}
\def \cem{8}
\def \cer{14}
\def \ced{5}
\def \c_32{1}
\def \down{3}

\foreach \s in {0,1,2,3}
{
  \node[circle,fill,radius = 0.2 pt , inner sep= 0.5pt,label=above:$a_\s$] at ({\ced+\radius*cos(3*360/\n * (\s ))}, {\ced+\radius*sin(3*360/\n * (\s ))} ) {$\s$};
}

\foreach \s in {0,2}
{
  \node[circle,fill,radius = 0.2 pt , inner sep= 0.5pt,label=above:$a_\s$] at ({\ced+\radius*cos(3*360/\n * (\s ))}, {\ced-\down+\radius*sin(3*360/\n * (\s ))} ) {$\s$};
}
\foreach \s in {1,3}
{
  \node[circle,fill,radius = 0.2 pt , inner sep= 0.5pt,label=right:$a_\s$] at ({\ced+5+\radius*cos(360/\n * (\s - 1))}, {\ced-\down+\radius*sin(360/\n * (\s - 1))} ) {$\s$};
}

\node[anchor=base] at (2, \ced )
                {$A_3$};
          \node[anchor=base] at (3, \ced )
                {$=$};
          \node[anchor=base] at (3, \ced-\down  )
                {$=$};                
\draw (\ced,\ced) circle (\radius);
\draw (\ced,\ced-\down) circle (\radius);
\draw (\ced+5,\ced-\down) circle (\radius);
%\draw (0,0) -- (4,0);

\node[anchor=base] at (7, \ced -\down)
                {$-$}; 
                
\node[anchor=base] at (8, \ced -\down)
                {$e^{i\theta}$};

                \end{tikzpicture} 
                
 So, you can get the FFT of $\{a_0,a_1,a_2,a_3\}$  using the following terms.
 
        
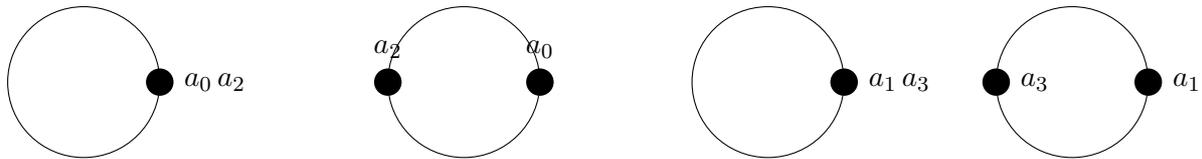
\begin{tikzpicture}

\def \n {4}
\def \radius {1}
\def \margin {8}
\def \cem{8}
\def \cer{14}
\def \ced{5}
\def \c_32{1}
\def \down{3}

\foreach \s in {0,2}
{
  \node[circle,fill,radius = 0.2 pt , inner sep= 0.5pt,label=above:$a_\s$] at ({\ced+\radius*cos(3*360/\n * (\s ))}, {\ced-\down+\radius*sin(3*360/\n * (\s ))} ) {$\s$};
}
\foreach \s in {1,3}
{
  \node[circle,fill,radius = 0.2 pt , inner sep= 0.5pt,label=right:$a_\s$] at ({\ced+3+5+\radius*cos(360/\n * (\s - 1))}, {\ced-\down+\radius*sin(360/\n * (\s - 1))} ) {$\s$};
}

\draw (\ced,\ced-\down) circle (\radius);
\draw (\ced+5+3,\ced-\down) circle (\radius);
%\draw (0,0) -- (4,0);

\foreach \s in {0}
{
  \node[circle,fill,radius = 0.2 pt , inner sep= 0.5pt,label=right:$a_\s\,a_2$] at ({\ced-5+\radius*cos(2*360/\n * (\s ))}, {\ced-\down+\radius*sin(2*360/\n * (\s ))} ) {$\s$};
}
\foreach \s in {1}
{
  \node[circle,fill,radius = 0.2 pt , inner sep= 0.5pt,label=right:$a_\s\,a_3$] at ({\ced-5+9+\radius*cos(360/\n * (\s - 1))}, {\ced-\down+\radius*sin(360/\n * (\s - 1))} ) {$\s$};
}

\draw (\ced-5,\ced-\down) circle (\radius);
\draw (\ced-5+9,\ced-\down) circle (\radius);

                \end{tikzpicture} 
 
 The first two terms are the FFT of  $\{a_0,a_2\}$ and the last two form the FFT of  $\{a_1,a_3\}$. Thus, an FFT can be decomposed into an FFT of the even terms and one of the odd terms. This saves a lot of computational cost since doing DFT naively is $\mathcal{O}(N^2)$. This is different from decomposing sums which had a cost of $\mathcal{O}(N)$ and where decomposing did not help save computational cost.
       
$\text{FFT}\{a_1,a_2,a_3,a_4\}$ has the combined computational cost of  $\text{FFT}\{a_1,a_3\}$ , $\text{FFT}\{a_2,a_4\}$ and $cN$ where c represents the cost of the multiplication and addition for each $A_n$. 

Expanding the FFTs recursively, $\text{FFT}\{a_1,a_2,a_3,a_4\}$ has the combined computational cost of  $\text{FFT}\{a_1\},\text{FFT}\{a_3\}$ , $\text{FFT}\{a_2\},\text{FFT}\{a_4\}$ and $2cN$. 

Computing the DFT naively with $N$ points requires $c_1N^2$ work while computing two DFTs of size $\frac{N}{2}$ requires only half as much work, $2 c_1 \big(\frac{N}{2}\big)^2 = \frac{1}{2} c_1 N^2$, plus $c_2 N$ work to combine the two results.

We can apply this recursively $\log_2 N$ times to completely eliminate the quadratic cost, leaving only the cost of $N$ 1-point DFTs plus $\log_2 N$ combinations each requiring $\mathcal{O}(N)$ work, for a total of $\mathcal{O}(N \log_2 N)$ work.

Thus, in general, the total cost is $N + c log_2(N)N \approx \mathcal{O}(log_2(N)N) $

\section*{Acknowledgments}
The author thanks David I. Ketcheson and Randall J. LeVeque  for their comments on earlier drafts.

\end{document}